\documentclass[preprintnumbers,floats,twocolumn,prd,aps]{revtex4}
\usepackage{amssymb,amsmath}

\newcommand{\K}{{\cal K}} \newcommand{\D}{{\cal D}}

\begin{document}

\title{Can Inflating Braneworlds be Stabilized?}
\author{Andrei V. Frolov}\email{frolov@cita.utoronto.ca}
\author{Lev Kofman}\email{kofman@cita.utoronto.ca}
\affiliation{
  CITA, University of Toronto\\
  Toronto, ON, Canada, M5S 3H8
}
\date{\today}

\begin{abstract}
  We investigate scalar perturbations from inflation
  in braneworld cosmologies with extra dimensions. For this we
  calculate scalar metric fluctuations around five dimensional warped
  geometry with four dimensional de~Sitter slices. The background
  metric is determined self-consistently by the (arbitrary) bulk scalar
  field potential, supplemented by the boundary conditions at both
  orbifold branes. Assuming that the inflating branes are stabilized
  (by the brane scalar field potentials), we estimate the lowest
  eigenvalue of the scalar fluctuations -- the radion mass. In the
  limit of flat branes, we reproduce well known estimates of the
  positive radion mass for stabilized branes. Surprisingly, however, we
  found that for de~Sitter (inflating) branes the square of the radion
  mass is typically negative, which leads to a strong tachyonic
  instability. Thus, parameters of stabilized inflating braneworlds
  must be constrained to avoid this tachyonic instability. Instability
  of ``stabilized'' de~Sitter branes is confirmed by the {\tt
  BraneCode} numerical calculations in the accompanying paper
  \cite{branecode}. If the model's parameters are such that the radion
  mass is smaller than the Hubble parameter, we encounter a new
  mechanism of generation of primordial scalar fluctuations, which have
  a scale free spectrum and acceptable amplitude.
\end{abstract}

\preprint{CITA-2003-29}
\maketitle

\section{Introduction}

One of the most interesting recent developments in high energy physics
has been the picture of braneworlds. Higher dimensional formulations
of braneworld models in superstring/M theory, supergravity and
phenomenological models of the mass hierarchy have the most obvious
relevance to cosmology. In application to the very early universe this
leads to braneworld cosmology, where our 3+1 dimensional universe is a
3d curved brane embedded in a higher-dimensional bulk \cite{review}.
Early universe inflation in this picture corresponds to 3+1 (quasi)
de~Sitter brane geometry, so that the background geometry is simply
described by the five dimensional warped metric with four dimensional
de~Sitter slices
\begin{equation}\label{warp}
  ds^2 = a^2(w)\left[dw^2 - dt^2 + e^{2Ht}d\vec{x}^2\right].
\end{equation}
For simplicity we use spatially flat slicing of the de~Sitter metric
$ds^2_4$. The conformal warp factor $a(w)$ is determined self-consistently
by the five-dimensional Einstein equations, supplemented by the
boundary conditions at two orbifold branes. We assume the presence of a
single bulk scalar field $\varphi$ with the potential $V(\varphi)$ and
self-interaction potentials $U_\pm(\varphi)$ at the branes. The
potentials can be pretty much arbitrary as long as the phenomenology of
the braneworld is acceptable. The class of metrics (\ref{warp}) with
bulk scalars and two orbifold branes covers many interesting braneworld
scenarios including the Ho\v{r}ava-Witten theory \cite{HW,Lukas}, the
Randall-Sundrum model \cite{RS1,RS2} with phenomenological
stabilization of branes \cite{GW,Dewolfe}, supergravity with domain
walls, and others \cite{FTW,FFK}.

We will consider models where by the choice of the bulk/brane
potentials the inter-brane separation (the so-called radion) can be fixed,
i.e. models in which branes could in principle be stabilized. The
theory of scalar fluctuations around flat stabilized branes, involving
bulk scalar field fluctuations $\delta\varphi$, scalar 5d metric
fluctuations and brane displacements, is well understood \cite{Tanaka:2000er}.
Similar to Kaluza-Klein (KK) theories, the extra-dimensional dependence can
be separated out, and the problem is reduced to finding the eigenvalues
of a second-order differential equation for the extra-dimensional
($w$-dependent) part of the fluctuation eigenfunctions subject to the
boundary conditions at the branes. The lowest eigenvalue corresponds to
the radion mass, which is positive $m^2>0$ and exceeds the TeV scale or so
\cite{Csaki:1999mp}. Tensor fluctuations around flat stabilized branes
are also stable.

Brane inflation, like all inflationary models, generates long
wavelength cosmological perturbations from the vacuum fluctuations of
all light (i.e. with mass less than the Hubble parameter $H$) degrees
of freedom. The theory of metric fluctuations around the background
geometry (\ref{warp}) with inflating (de~Sitter) branes is more
complicated than that for the flat branes. For tensor fluctuations
(gravitational waves), the lowest eigenvalue of the extra dimensional
part of the tensor eigenfunction is zero, $m=0$, which corresponds to
the usual 4d graviton. As it was shown in \cite{LMW,gw}, massive KK
gravitons have a gap in the spectrum; the universal lower bound on the
mass is $m \ge \sqrt{3 \over 2}\, H$. This means that massive KK tensor
modes are not generated from brane inflation. Massless scalar and
vector projections of the bulk gravitons are absent, so only the
massless 4d tensor mode is generated.

Scalar cosmological fluctuations from inflation in the braneworld
setting (\ref{warp}) have been considered in many important works
\cite{Mukohyama:2000ui, Kodama:2000fa, Langlois:2000ia,
vandeBruck:2000ju, Koyama:2000cc, Deruelle:2000yj, Gen:2000nu,
Mukohyama:2001ks}. The theory of scalar perturbations in braneworld
inflation with bulk scalars is even more complicated than for tensor
perturbations. This is because one has to consider 5d scalar metric
fluctuations and brane displacements induced not only by the bulk
scalar field fluctuations $\delta\varphi$, but also by the fluctuations
$\delta \chi$ of the inflaton scalar field $\chi$ living at the brane.
In fact, most papers on scalar perturbations from brane inflation
concentrated mainly on the inflaton fluctuations $\delta \chi$, while
the bulk scalar fluctuations were not included. This was partly because
in the earlier papers on brane inflation people considered a single
brane embedded in an AdS background without a bulk scalar field, and
partly because for braneworlds with two stabilized branes there was an
expectation that the fluctuations of the bulk scalar would be massive
and thus would not be excited during inflation.

In this letter we focus on the bulk scalar field fluctuations, assuming
for the sake of simplicity that the inflaton fluctuations $\delta \chi$
are subdominant. We consider a relatively simple problem of scalar
fluctuations around curved (de~Sitter) branes, involving only bulk
scalar field fluctuations $\delta\varphi$. We find the
extra-dimensional eigenvalues of the scalar fluctuations subject to
boundary conditions at the branes, focusing especially on the radion
mass $m^2$ for the inflating branes. In particular, we investigate the
presence or absence of a gap in the KK spectrum of scalar fluctuations
in view of the tensor mode result. Our results are a generalization of
the known results for flat stabilized branes \cite{Tanaka:2000er},
which we reproduce in the limit where the branes are flattening $H \to 0$.

\section{Bulk Equations}

The five-dimensional braneworld models with a scalar field in the bulk
are described by the action
\begin{eqnarray}\label{eq:action}
S &=& M_5^3 \int \sqrt{-g}\, d^5 x\,
           \left\{R - (\nabla\varphi)^2 - 2V(\varphi)\right\} \nonumber\\
  && -2 M_5^3 \sum \int \sqrt{-q}\, d^4 x\,
           \left\{ [\K] + U(\varphi)\right\},
\end{eqnarray}
where the first term corresponds to the bulk and the sum contains
contributions from each brane. The jump of the extrinsic curvature
$[{\cal K}]$ provides the junction conditions across the branes (see
equation (\ref{eq:jc}) below). Variation of this action gives the bulk
Einstein $G_{AB}=T_{AB}(\varphi)$ and scalar field $\Box\varphi=V_{,\varphi}$
equations. For the (stationary) warped geometry (\ref{warp}) they are
\begin{subequations}\label{eq:bg}
\begin{eqnarray}
  &\displaystyle \varphi'' + 3\frac{a'}{a} \varphi' - a^2 V' = 0,&\label{eq:bg:phi}\\
  &\displaystyle \frac{a''}{a} = 2\, \frac{a'^2}{a^2} - H^2 - \frac{\varphi'^2}{3},&\label{eq:bg:a}\\
  &\displaystyle 6\left(\frac{a'^2}{a^2} - H^2\right) = \frac{\varphi'^2}{2} - a^2 V,&\label{eq:bg:c}
\end{eqnarray}
\end{subequations}
where the prime denotes the derivative with respect to the extra
dimension coordinate $w$. The first two equations are dynamical, and
the last is a constraint. The solutions of equations (\ref{eq:bg}) were
investigated in detail in \cite{FFK}.

Now we consider scalar fluctuations around the background (\ref{warp}).
The perturbed metric can be written in the longitudinal gauge as
\begin{equation}\label{eq:metric:pert}
  ds^2 = a(w)^2 \left[(1+2\Phi) dw^2 + (1+2\Psi)ds_4^2\right].
\end{equation}
The linearized bulk Einstein equations and scalar field equation relate
two gravitational potentials $\Phi(x^A)$, $\Psi(x^A)$ and bulk scalar
field fluctuations $\delta\varphi(x^A)$. The off-diagonal Einstein
equations require that
\begin{equation}
  \Psi = - \frac{\Phi}{2},
\end{equation}
similar to four-dimensional cosmology, although the coefficient is
different.

The symmetry of the background guarantees separation of variables, so
that perturbations can be decomposed with respect to four-dimensional
scalar harmonics, e.g.
\begin{equation}\label{eq:sep}
  \Phi(x^A) = \sum\limits_m \Phi_m(w) Q_m(t, \vec x),
\end{equation}
where the eigenvalues $m$ (constant of separation) appear as the
four-dimensional masses ${^4}\Box Q_m = m^2 Q_m$, where ${^4}\Box$ is
the D'Alembert operator on the 4d de~Sitter slice. The four-dimensional
massive scalar harmonics $Q_m$ can be further decomposed as
$Q_m(t,\vec x) = \int f_k^{(m)}(t)\, e^{i \vec k \vec x}\, d^3k$.
The temporal mode functions $f_k^{(m)}(t)$ obey the equation
\begin{equation}\label{eq:4}
  \ddot{f} + 3H\dot{f} + \left( e^{-2Ht}k^2 + m^2 \right) f = 0,
\end{equation}
where dot denotes time derivative, and we dropped the labels $k$ and
$m$ for brevity.

Out of the remaining linearized Einstein equation we get the following
equations for the extra-dimensional mode functions $\Phi_m(w)$ and
$\delta\varphi_m(w)$
\begin{subequations}\label{eq:pert}
\begin{eqnarray}
  (a^2 \Phi)' &=& \frac{2}{3} a^2 \varphi'\, \delta\varphi,\\
  \left(\frac{a}{\varphi'}\, \delta\varphi\right)' &=&
    \left(1 - \frac{3}{2} \frac{m^2+4H^2}{\varphi'^2}\right) a \Phi,
\end{eqnarray}
\end{subequations}
where we again omitted the label $m$ for transparency.

These are very similar to the scalar perturbation equations in
four-dimensional cosmology with a scalar field \cite{mukhanov}, except
for some numerical coefficients and powers of $a(w)$ (because the
spacetime dimensionality is higher), and up to time to extra spatial
dimension exchange. Indeed, we can introduce the higher-dimensional
analog of the Mukhanov's variable. However, in the presence of the
curvature term $H^2$, the eigenvalue $m^2$ enters the second order
equation for it in a complicated way, similar to that in the 4d problem
with non-zero spatial curvature, see e.g.~\cite{Garriga:1999vw}.
We can introduce another convenient variable 
$u_m = \sqrt{\frac{3}{2}}\frac{a^{3/2}}{\varphi'}\, \Phi_m$.
Then the two first order differential equations (\ref{eq:pert}) can be
combined into a single Schr\"odinger-type equation
\begin{equation}\label{v}
  u_m'' + \Big( m^2+4H^2 - V_{\text{eff}}(w) \Big) u_m = 0
\end{equation}
with the effective potential $V_{\text{eff}} = \frac{z''}{z} +
\frac{2}{3}\varphi'^2$, where we defined $z = \left(\frac{2}{3} a
\varphi'^2\right)^{-\frac{1}{2}}$.

There are two main differences relative to the four dimensional
cosmology. First, in the latter case, FRW geometry with {\it flat} 3d
spatial slices is usually considered, while the five dimensional brane
inflation metric has {\it curved} 4d slices, which results in extra
terms like $4H^2$ in equation (\ref{v}). Second, here we are dealing
not with an \emph{initial} but a \emph{boundary} value problem, with
associated boundary conditions for perturbations at the branes on the
edges. After we derive the boundary conditions, we will calculate the
KK spectrum of the eigenvalues $m$.

\section{Brane Embedding and Boundary Conditions}

The embedding of each brane is described by $w=w_{\pm}+\xi_\pm(x^a)$,
where $\xi_\pm$ is the transverse displacement of the perturbed brane
and $w_\pm$ is the position of the unperturbed brane. Holonomic basis
vectors along the brane surface are
  $e_{(a)}^A \equiv \frac{\partial x^A}{\partial x^a}
   = \Big(\xi_{,a}, \delta_a^A\Big)$,
while the unit normal to the brane is
  $n_{A} = a \Big(1+\Phi, -\xi_{,a} \delta_A^a\Big)$.
The induced four-metric on the brane $d\sigma^2 = q_{ab} dx^a dx^b$ does
not feel the brane displacement (to linear order) and is conformally
flat
\begin{equation}\label{eq:induced}
  d\sigma^2 = a^2(1-\Phi)\left[-dt^2 + e^{2Ht}d\vec{x}^2\right].
\end{equation}
The junction conditions for the metric and the scalar field at the brane are
\begin{equation}\label{eq:jc}
  [\K_{ab} - \K q_{ab}] = U(\varphi) q_{ab}, \hspace{1em}
  [n\cdot\nabla\varphi] = \frac{\partial U}{\partial \varphi},
\end{equation}
where the extrinsic curvature is defined by $\K_{ab} = e_{(a)}^A
e_{(b)}^B n_{A;B}$. We will only need its trace, which up to linear
order in perturbations is
\begin{equation}\label{eq:k}
  \K = 4\frac{a'}{a^2} - 2 \frac{(a^2\Phi)'}{a^3} - \frac{{^4}\Box\xi}{a}.
\end{equation}

For the background geometry (under the assumption of reflection
symmetry across the branes), equations (\ref{eq:jc}) reduce to
\begin{equation}\label{eq:jc:bg}
  \frac{a'}{a^2} = \mp \frac{U}{6}, \hspace{1em}
  \frac{\varphi'}{a} = \pm \frac{U'}{2}.
\end{equation}
For the perturbed geometry, the traceless part of the extrinsic
curvature must vanish in the absence of matter perturbations on the
brane. Since it contains second cross-derivatives of $\xi$, the brane
displacement $\xi$ is severely restricted. Basically, this means that
the oscillatory modes of brane displacement are not excited without
matter support at the brane. While there could possibly be global
displacements of the brane, they do not interest us, so in the
following we set $\xi=0$. Of course, for the more complete problem
which includes fluctuations $\delta \chi$ of the ``inflaton'' field on
the brane, the displacement $\xi$ does not vanish.

Using expression (\ref{eq:k}) for the trace of the extrinsic curvature,
the first of equations (\ref{eq:jc}) gives us the junction condition
for linearized perturbations at the two branes
  $(a^2 \Phi)'\big|_{w_{\pm}} =
    \pm \frac{1}{3}\, U' a^3\, \delta\varphi \big|_{w_{\pm}}$.
However, this junction condition does not really place any further
restrictions on the bulk field perturbations, as it identically follows
from the bulk perturbation equations (\ref{eq:pert}) and the background
junction condition (\ref{eq:jc:bg}). Rather, this junction condition
would relate the brane displacement $\xi$ to the matter perturbations
on the brane if they were not absent.

The second of equations (\ref{eq:jc}) gives us a physically relevant
boundary condition for the bulk field perturbations
\begin{equation}
  (\delta\varphi' - \varphi' \Phi)\big|_{w_{\pm}} =
    \pm \frac{1}{2}\, U'' a\, \delta\varphi \big|_{w_{\pm}}.
\end{equation}
Using the bulk equations (\ref{eq:pert}), this can be rewritten in a
more suggestive form
\begin{equation}\label{eq:bc}
  \left(\frac{a}{\varphi'}\, \delta\varphi\right)\Bigg|_{w_{\pm}} =
    \frac{3}{2} \frac{m^2+4H^2}{a\varphi'^2}
    \frac{a^2 \Phi}{\frac{a^2 V'}{\varphi'} - 4 \frac{a'}{a} \mp a U_\pm''} \Bigg|_{w_{\pm}}.
\end{equation}
The eigenvalues $m^2$ of bulk perturbation equations subject to the
boundary condition (\ref{eq:bc}) form a KK spectrum, which we find
numerically. We considered several examples of the potentials $V$ and
$U_{\pm}$, and found no universal positive mass gap. Moreover, for the
most interesting models we found negative $m^2$.

To understand the KK spectrum of $m^2$, we make a simplification of the
boundary condition (\ref{eq:bc}) which will allow us to treat the
eigenvalue problem analytically, and which well corresponds to a spirit
of brane stabilization \cite{GW}. Indeed, \emph{rigid stabilization} of
branes is thought to be achieved by taking $U''$ (i.e. the brane mass
of the field) very large, so that the scalar field gets pinned down at
the positions of the branes. In this case, the right hand side of
(\ref{eq:bc}) becomes very small, which leads to the boundary condition
\begin{equation}\label{eq:stab}
  \delta\varphi\big|_{w_\pm} = 0.
\end{equation}
This by itself \emph{does not guarantee stability}, or vanishing of the
metric perturbations on the brane for that matter, as perturbations
live in the bulk and only need to satisfy (\ref{eq:stab}) on the
branes. This poses an eigenvalue problem for the mass spectrum of the
perturbation modes, which we study next.

\section{KK Mass Spectrum}

Unlike the situation with gravitational waves \cite{gw}, for the scalar
perturbations there is no zero mode with $m=0$, nor is there a
``supersymmetric'' factorized form of the ``Schr\"odinger''-like
equation (\ref{v}). To find the lowest mass eigenvalue, we have to use
other ideas. Powerful methods for analyzing eigenvalue problems
exist for normal self-adjoint systems \cite{kamke}. To use them, we
transform our eigenvalue problem (\ref{eq:pert}) and (\ref{eq:stab}) into the
self-adjoint form. While the second order differential equation
(\ref{v}) is self-adjoint, the boundary conditions for $u$ are not.
Therefore, we introduce a new variable $Y=u/z=a^2\Phi$ and impose the
boundary conditions (\ref{eq:stab}) to obtain the boundary value
problem
\begin{subequations}\label{eq:evp}
\begin{eqnarray}
  \label{eq:evp:de} &\D Y \equiv -(gY')' + fY = \lambda gY,&\\
  \label{eq:evp:bc} &Y'(w_{\pm}) = 0,&
\end{eqnarray}
\end{subequations}
where we have introduced the short-hand notation $f = 1/a$, $g = z^2 =
\left(\frac{2}{3} a \varphi'^2\right)^{-1}$, and $\lambda = m^2+4H^2$.
Since the boundary value problem (\ref{eq:evp}) is self-adjoint, it is
guaranteed that the eigenvalues $\lambda$ are real and non-negative,
$\lambda \ge 0$. To estimate the lowest eigenvalue $\lambda_1$ of the
eigenvalue problem (\ref{eq:evp}), we apply the Rayleigh's formula
\cite{kamke}, which places a rigorous upper bound on $\lambda_1$
\begin{equation}
  \lambda_1 \le \frac{\int F \D F\, dw}{\int g F^2\, dw},
\end{equation}
where $F$ can be \emph{any} function satisfying the boundary conditions
(\ref{eq:evp:bc}), and does not have to be a solution of
(\ref{eq:evp:de}). Taking a trial function $F=1$, we have
\begin{equation}
  \lambda_1 \le \frac{\int f\, dw}{\int g\, dw}.
\end{equation}
This bound on the lowest mass eigenvalue is our main result:
\begin{equation}\label{eq:bound}
  m^2 \le -4H^2 + \frac{2}{3} \frac{\int \frac{dw}{a}}{\int \frac{dw}{a\varphi'^2}}.
\end{equation}
In practice, $F=1$ is a pretty good guess for the lowest eigenfunction,
so the bound (\ref{eq:bound}) is usually close to saturation (up to a
few percent accuracy in some cases), as we have observed in direct
computations using a numerical eigenvalue finder.

The right hand side of equation (\ref{eq:bound}) has the structure
$-4H^2 + m_0^2(H)$, where the second term is a functional of $H$
(including the implicit $H$-dependence of the warp factor $a$). In the
limit of flat branes $H \to 0$ we have only the second, positive term.
In this limit our expression agrees with the estimation of the radion
mass $m_0^2$ for flat branes, obtained in various approximations
\cite{Csaki:1999mp,Tanaka:2000er,Mukohyama:2001ks}. A non-vanishing $H$
alters $m^2$ through both terms. The most drastic alteration of $m^2$
due to $H$ comes from the big negative term $ -4H^2$. For the
particular case of two de~Sitter branes embedded in 5d AdS without a
bulk scalar this negative term was noticed in \cite{Gen:2000nu}.

\section{Tachyonic Instability of the Radion for Inflating Branes}

The most striking feature of the mass bound (\ref{eq:bound}) is that
$m^2$ for de~Sitter branes is typically negative. Trying, for instance,
to do Goldberger-Wise stabilization of braneworlds with inflating
branes while taking bulk gradients $\varphi'^2$ small enough to ignore
their backreaction (as it is commonly done for flat branes) is a sure
way to get a tachyonic radion mass: an estimate of the integrals gives
$m^2 \le -4H^2 + O(\varphi'^2)$, which will go negative if the bulk
scalar field is negligible $\varphi'^2 \ll H^2$.

In what follows we consider two situations. In this section, we
consider braneworld models where $m^2$ is negative and mostly defined
by the first term $-4H^2$ in equation (\ref{eq:bound}). In the next
section, we consider the case where both terms in equation
(\ref{eq:bound}) are tuned to be comparable and the net radion mass is
smaller than the Hubble parameter $|m^2| \leq H^2$. In the last section
we will discuss how these two cases may be dynamically connected.

Suppose we start with a braneworld with curved de~Sitter branes, and we
find the mass squared of the radion to be negative. The
extra-dimensional eigenfunction $\Phi_m(w)$ is regular in the interval
$w_{-} \leq w \leq w_{+}$. Let us turn, however, to the
four-dimensional eigenfunction $Q_m(t, \vec x)$. Bearing in mind the
evolution of the quantum fluctuations of the bulk field, we choose the
positive frequency vacuum-like initial conditions in the far past $t
\to -\infty $, $f_k(t) \simeq \frac{1}{\sqrt{2k}} e^{ik\eta}$,
$\eta=\int dt\, e^{-Ht}$. For the tachyonic mode $m^2 <0$ the solution
to equation (\ref{eq:4}) with this initial condition is given in terms
of Hankel functions $f_k^{(m)}(\eta)=\frac{\sqrt{\pi}}{2} H
|\eta|^{3/2} {\cal H}^{(1)}_{\mu} (k\eta)$, with the index
$\mu=\sqrt{\frac{9}{4}+\frac{|m^2|}{H^2}}$. The late-time asymptotic of
this solution diverges exponentially as $t \to \infty$ ($\eta \to 0$)
\begin{equation}\label{asym}
  f_k^{(m)}(t) \propto \exp \left[\left( {\sqrt{\frac{9}{4}+\frac{|m^2|}{H^2}} - \frac{3}{2}} \right) H t \right].
\end{equation}
Thus the negative tachyon mass of the radion $|m^2| \sim 4H^2$ leads to
a strong exponential instability of scalar fluctuations $\Phi \propto
e^{Ht}$. This instability is observed using a completely different
method in the accompanying {\tt BraneCode} paper \cite{branecode},
where we give a fully non-linear numerical treatment of inflating
branes which were initially set to be stationary by the potentials
$U_{\pm}(\varphi)$, and without any simplifications like approximating
boundary condition (\ref{eq:bc}) with (\ref{eq:stab}).

Tachyonic instability of the radion for inflating branes means that, in
general, {\it braneworlds with inflation are hard to stabilize.} From
the point of view of 4d effective theory one would expect brane
stabilization at energies lower than the mass of the flat brane radion
$m_0^2$, which is roughly equal to the second term in (\ref{eq:bound}).
If the energy scale of inflation $H$ is larger than $m_0$, $H^2 \gg
m_0^2$, this expectation is incorrect.

Successful inflation (lasting more than $65 H^{-1}$) requires the
radion mass $m^2$ to be not too negative
\begin{equation}\label{criter}
  m^2 \gtrsim - \frac{H^2}{20}.
\end{equation}
This is possible if both terms in (\ref{eq:bound}) are of the same
order. In the popular braneworld models the radion mass in the low
energy limit, $m_0$, is of order of a TeV. For these models the scale
of ``stable'' inflation would be the same order of magnitude, $H \sim
\text{TeV}$. Although there is no evidence that this scale of inflation
is too low, it is not a comfortable scale from the point of view of the
theory of primordial perturbations from inflation.

It is interesting to note that the system of curved branes may
dynamically re-configure itself to reach a state where the condition
(\ref{criter}) is satisfied. In the case of the bulk scalar field
$\varphi$ acting alone, for quadratic potentials $U_{\pm}$ suitable for
brane stabilization, there may be two stationary warped geometry
solutions (\ref{warp}) with two different values of $H$. The solution
with the larger Hubble parameter $H$ might be dynamically unstable due
to the tachyonic instability of the radion, which we described above.
The second solution with the lower $H$ which satisfies (\ref{criter})
might be stable. A fully non-linear study of this model was performed
numerically with the {\tt BraneCode} and is reported in the
accompanying paper \cite{branecode}. It shows that, indeed, the
tachyonic instability violently re-configures the starting brane state
with the larger $H$ into the stable brane state with the lower $H$.
This re-configuration of the brane system has a spirit of the Higgs
mechanism.

If we add an ``inflaton'' scalar field $\chi$ located at the brane,
its slow roll contributes to the decrease of $H$.

Thus, for the ``stable'' brane we have a radion mass (\ref{criter}).
This condition includes the case when the radion is lighter than
$H$, $|m^2| < H^2$. Even if the radion tachyonic instability is
avoided, the light radion leads us to the other side of the story, a
new mechanism of generation of scalar fluctuations from inflation
associated with the radion.

\section{Induced Scalar Metric Perturbations at the Observable Brane}

Suppose that the radion mass is smaller than $H$, $|m^2| \ll H^2$, so
that from (\ref{asym}) we get the amplitude of the temporal mode
function $f_k^{(m)}(t)$ in the late time asymptotic frozen at the level
$f_k^{(m)}(t) \simeq \frac{H}{\sqrt{2}k^{3/2}}$. This is nothing but
the familiar generation of inhomogeneities of a light scalar field from
its quantum fluctuations during inflation. Therefore an observer at the
observable brane will encounter long wavelength scalar metric
fluctuations generated from braneworld inflation.

The four dimensional metric describing scalar fluctuations around an
inflating background is usually written as
\begin{equation}\label{eq:induced1}
  d\sigma^2 = - (1+2\widetilde{\Phi}) d\tilde{t}^2
    + (1-2\widetilde{\Psi})e^{2\widetilde{H}\tilde{t}}d\tilde{x}^2,
\end{equation}
where $\widetilde{\Phi}$ and $\widetilde{\Psi}$ are scalar metric
fluctuations. The induced four-metric on the brane (\ref{eq:induced})
in our problem can be rewritten in this standard form (\ref{eq:induced1})
if we absorb the (constant) warp factor $a(w_{+})$ in the redefined time
$\tilde{t} = at$ and spatial coordinates $\tilde{x} = a\vec{x}$ and
rescale the Hubble parameter $\widetilde{H} = H/a$. Then we see that the
induced scalar perturbations on the brane are
\begin{equation}\label{prop}
  \widetilde{\Psi} = -\widetilde{\Phi} = \frac{1}{2}\, \Phi.
\end{equation}
The sign of the first equality here is opposite to what we usually have
for $3+1$ dimensional inflation with a scalar field. It implies that
the 4d Weyl tensor of the induced metric vanishes, as the induced
fluctuations are conformally flat. The conformal structure of
fluctuations (\ref{prop}) is typical \cite{km87} for a $R^2$ inflation
in the Starobinsky model \cite{star}. It is not a surprise, because for
the scale of inflation comparable to the mass $m_0$ of the flat brane
radion we expect higher derivative corrections to the 4d effective
gravity on the brane. Indeed, the massive radion corresponds to a
higher derivative 4d gravity \cite{Mukohyama:2001ks}.

The amplitude and spectrum of induced fluctuations is defined by
$\Phi$. From the mode decomposition (\ref{eq:sep}) we get
\begin{equation}\label{ampl}
  k^{3/2}\, \widetilde{\Phi}_k \simeq \Phi_m (w_+) \, \frac{H}{M_4},
\end{equation}
where $\Phi_m(w_+)$ is the amplitude of the extra-dimensional eigenmode
at the observable brane, normalized in such a way that the fluctuations
$\Phi(w, t, {\vec x})$ are canonically quantized on the 4d slice, namely
$M_5^3 \int \frac{3}{2} \frac{a^3}{\varphi'^2}\, |\Phi_m(w)|^2 \, dw = 1$.
The normalization $M_4$ of the 4d mode functions follows from canonical
quantization of the perturbed action (\ref{eq:action}); the usual 4d
Planck mass $M_p$ is expected to be recovered in the effective field
theory on the observable brane \cite{Tanaka:2000er}.

The scalar metric fluctuations induced by the bulk scalar field
fluctuations are scale-free and have the amplitude $k^{3/2}
\widetilde{\Phi}_k \propto \frac{H}{M_p}$, with the numerical coefficient
depending on the details of the warped geometry. The nature of these
fluctuations is very different from those in $(3+1)$-dimensional
inflation, where the inflaton scalar field is time dependent. Induced
scalar fluctuations do not require ``slow-roll'' properties of the
potentials $V$ and $U_{\pm}$. The underlying background bulk scalar
field has no time-dependence, but only $y$ dependence. Thus, generation
of induced scalar metric fluctuations from braneworld inflation is a
new mechanism for producing cosmological inhomogeneities.

If we add another, inflaton field $\chi$ localized at the brane, we
should expect that fluctuations of both fields, the bulk scalar $\delta
\varphi$ and the inflaton $\delta \chi$, contribute to the metric
perturbations. We can conjecture that the net fluctuations will be
similar to those derived in the combined model with $R^2$ gravity and a
scalar field \cite{kls}.

\section{Discussion}

Let us discuss the physical interpretation and the meaning of our
result. Stabilization of flat branes is based on the balance between
the gradient $\phi'$ of the bulk scalar field and the brane potentials
$U(\phi)$ which keeps $\phi$ pinned down to its values $\phi_i$ at the
branes. The interplay between different forces becomes more delicate if
the branes are curved. The warped configuration of curved branes has
the lowest eigenvalue for scalar fluctuations around it
\begin{equation}\label{crit}
  m^2=-4H^2+m_0^2(H) \ .
\end{equation}
The term $m_0^2(H)$ is a functional of $H$, and depends on the parameters
of the model. If parameters are such that $m^2$ becomes negative due to
excessive curvature $\sim H^2$, the brane configuration becomes
unstable. This is analogous to an instability of a simple elastic
mechanical system supported by the balance of opposite forces, which
arises for a certain range of the underlying parameters.

Tachyonic instability of curved branes has serious implications for
the theory of inflation in braneworlds. It may be not so easy to have a
realization of inflation in the braneworld picture without taking care
of parameters of the model. Inflation where $m^2$ in (\ref{crit}) is
negative and $|m^2|$ is larger than $H^2$ is a short-lived stage
because of this instability. After inflation, the late time evolution
should bring the brane configuration to (almost) flat stabilized branes
in the low energy limit. This by itself requires fine tuning of the
potentials $V$ and $U_{\pm}$ to provide stabilization. Stabilization at
the inflation energy scale requires extra fine tuning to get rid of the
tachyonic effect.

Working with a single bulk scalar field, it is probably not easy to
simultaneously achieve stabilization not only at low energy, but also
at the high energy scale of inflation, to insure that $|m^2| \ll H^2$,
and to provide a graceful exit from inflation. One may expect that
introduction of another scalar field $\chi$ on the brane can help to
have stabilization both at the scale of inflation and in the low energy
limit. If we can achieve brane stabilization during inflation by
suppression of the tachyonic instability, we encounter a byproduct
effect. Light modes of radion fluctuations inevitably contribute to the
induced scalar metric perturbations. Therefore the theory of braneworld
inflation has an additional mechanism of generation of primordial
cosmological perturbations. This new mechanism is different from that
of the usual 4d slow roll inflation.

It appears that one of the most interesting potential applications of
our effect is a mechanism for reducing the 4d effective cosmological
constant at the brane. Indeed, in terms of brane geometry, the 4d
cosmological constant is related to the 4d curvature of the brane.
Suppose we have two solutions of the background equations (\ref{eq:bg})
with higher and lower values of the curvature of de Sitter brane,
which is proportional to $H^2$. (The existence of two solutions for certain
choices of parameters of the Goldberger-Wise type potentials used for
brane stabilization can be demonstrated, see \cite{FFK,branecode}.)
Suppose that the solution with the larger value of brane curvature is
unstable. Then the brane configuration will violently restructure
into the other static configuration, which is characterized by the
lower value of brane curvature where the tachyonic instability is absent. The
branes are flattening, which for a 4d observer means the lowering of the
cosmological constant. It will be interesting to investigate how this
mechanism works for brane configurations with several scalar fields or
potentials which can admit more than two static solutions.

The problem of the cosmological constant from a braneworld perspective
(as a flat brane) was discussed in the literature. There was a
suggestion that the flat brane is a special solution of the bulk
gravity/dilaton system with a single brane \cite{Arkani-Hamed:2000eg,
Kachru:2000hf}, the claim which was later dismissed \cite{Forste:2000ft}.
In our setup, we consider two branes in order to screen the naked
bulk singularity, which was one of the factors spoiling the models
\cite{Arkani-Hamed:2000eg, Kachru:2000hf}. The new element which
emerges from our study is the instability of the curved branes.

\section*{Acknowledgments}

We are grateful to R. Brandenberger, J. Cline, C. Deffayet, J. Garriga,
A. Linde, S. Mukohyama, D. Pogosyan and V. Rubakov for valuable
discussions. We are especially indebted to our collaborators on the
{\tt BraneCode} project, G. Felder, J. Martin and M. Peloso. This
research was supported in part by the Natural Sciences and Engineering
Research Council of Canada and CIAR.


\end{document}